\documentclass[12pt]{article}
\usepackage{amssymb}

\usepackage{graphicx}
\usepackage{amsmath}
\usepackage{amsbsy}
\usepackage{color}
\font\tenscr=rsfs10  scaled 1200
\newcommand{\rf}[1]{(\ref{#1})}
\newcommand{\dst}{\displaystyle}
\newcommand{\B}{\boldsymbol}
\newcommand{\x}{\B x}
\newcommand{\y}{\B y}
\def\limfunc#1{\mathop{\rm #1}}%

\newcommand{\bk}{\B k}
\newcommand{\bt}{\widehat{\x}}
\newcommand{\supp}{\limfunc{supp}}
\newcommand{\R}{\mathbb {R}}

\newcommand{\Sp}{\mathbb {S}}

\newcommand{\K}{\mbox {\tenscr K}}
\newcommand{\ta}{\widehat{\x}}

\renewcommand{\leq}{\leqslant}
\renewcommand{\geq}{\geqslant}
\renewcommand{\phi}{\varphi}
\renewcommand{\kappa}{\varkappa}

\newcount\minutes
\newcount\scratch

\def\timestamp{%
\scratch=\time
\divide\scratch by 60
\edef\hours{\the\scratch}
\multiply\scratch by 60
\minutes=\time
\advance\minutes by -\scratch
\hours:\null
\ifnum\minutes< 10 0\fi
\the\minutes}

\title{On the determination of the boundary impedance\\ from the far field pattern}

\author{Yuri A. Godin\thanks{Department of Mathematics and Statistics, University of
North Carolina at Charlotte, Charlotte, NC 28223 ({\tt ygodin@uncc.edu, brvainbe@uncc.edu}).} \and Boris Vainberg$^\ast$}

\date{\timestamp \hspace{1mm} \today}

\begin{document}
\maketitle

\begin{abstract}
We consider the Helmholtz equation in the half space and
suggest two methods for determining the boundary impedance from knowledge of the far field pattern of the time-harmonic incident wave. We introduce a potential for which the far field patterns in specially selected directions represent its Fourier coefficients. The boundary impedance is then calculated from the potential by an explicit formula or from the WKB approximation. Numerical examples are  given to demonstrate efficiency of the approaches. We also discuss the validity of the WKB approximation in determining the impedance of an obstacle.
\end{abstract}

\markboth{\hfill \today \hfill}{\hfill \today \hfill}



\pagestyle{myheadings}
\thispagestyle{plain}
\markboth{Yu. A. GODIN and B. VAINBERG}{On the determination of the boundary impedance}

\section{Introduction}

%
Various impurities such as gases, dust,cracks, etc., on the surface
of a body subject to an incident wave can be modeled by the impedance boundary condition \cite{SM05}. The detection of these inhomogeneities using nondestructive testing is then reduced to the reconstruction of the
impedance from the measurements of scattering field \cite{H03}. 
Optical scanning of the surface of silicon wafers used for quality control in semiconductor industry \cite{BMF03} is one of the possible applications of this method. 

We consider the scattering of an incident time-harmonic plane wave from the boundary of the half-space $\R^3_{+} =\{{\x}=(x_1,x_2,x_3), \, x_3 >0\}$. The problem is described by the Helmholtz equation
\begin{equation}
-\Delta u = k^2 u, \quad x_3 > 0,
\label{r3}
\end{equation}
with the impedance boundary condition
\begin{equation}
\left. u_{x_3} + ik\gamma({\x^\prime})\, u \right|_{x_3 = 0} = 0, \quad {\x^\prime} = (x_1,x_2,0), \label{bcr3}
\end{equation}
where $\gamma ({\x^\prime})$ is the surface impedance with a bounded support $\supp \gamma \subset [-1,1]\times [-1,1]$
and $u=u({\x})$ is the superposition of the incident, reflected, and scattered waves
\begin{equation}
u({\x}) = e^{i{\bk}\cdot{\x}} + e^{i{\bk^{\ast}}\cdot{\x}}+ \psi({\x}).
\label{sup}
\end{equation}
Here ${\bk}=(k_{1},k_{2},k_{3})$ is a vector such that
$|{\bk}|=k$, ${\bk^\ast}=(k_{1},k_{2},-k_{3})$,
and function $\psi({\x})$ satisfies the radiation condition
\begin{equation}
\psi({\x}) = \frac{e^{ik|{\x}|}}{|{\x}|} \left[ f({\bk}, \ta) +
O\left( \frac{1}{|{\x}|} \right) \right], \quad {\ta} = \frac{{\x}}{|{\x}|} \in \Sp^2. \label{rc}
\end{equation}
The inverse scattering problem for \rf{r3}-\rf{bcr3} consists in determining the
 impedance $\gamma ({\x^\prime})$ by the far field pattern $f=f(\bk, \ta)$ when $\bk$ is fixed and $\ta \in \Sp^2$.

In the next section we introduce a modified potential $v$ and express the impedance $\gamma$ through $v$ using
an explicit formula. The mapping $v \to f$ is linear. Hence, the initial nonlinear inverse problem is split into two
steps: solution of a linear problem (restoring $v$ from $f$) and application the explicit formula. The similar approach was used in the discrete counterpart of the problem \cite{GV08}. This approach does not formally require $k \gg 1$.
We also modify it for large $k$  using the WKB method.

In the case of a bounded obstacle, the WKB method allows one to connect the impedance $\gamma$ with the asymptotic expansion of the far field pattern (see \cite{M76}). We perform a simple numerical calculations in order to find the range of parameters for which the WKB method can be used to determine the impedance. The inverse impedance problem has been considered in \cite{CK81}-\cite{CC04} for general bounded obstacles. 
Our assumptions simplify the problem, and as a result its analytical and numerical solutions become easier.

Note the difference in WKB approach in the inverse impedance problem for the half space and a bounded obstacle.
In the latter case the asymptotic expansion of the far field in a given direction is determined by the value of the impedance in a specific point if the obstacle is convex. This is not true for the half space.

\section{Explicit formula for the impedance}

We reduce the problem to the whole space $\R^3$ by extending function $u$ evenly through
the boundary $x_3 = 0$ for $x_3 < 0$. Then equations \rf{r3}-\rf{bcr3} are replaced by the Schr\"{o}dinger
equation
\begin{equation}
(-\Delta + q)\, u = k^2 u, \quad {\B x} \in \R^3,
\label{cont.s}
\end{equation}
where potential $q({\B x}) = -2ik\gamma ({\B x^\prime})\delta (x_3)$, and $\delta(x)$ is the Dirac delta-function.

Substituting \rf{sup} into \rf{cont.s}, we obtain that the scattering solution
$\psi(\x)$ satisfies the equation
\begin{equation}
(-\Delta + q(\x)-k^2) \psi = -q(\x) \left(e^{i\bk \cdot \x} + e^{i\bk^\ast
\cdot \x}\right)=-2q(\x) e^{i\bk^\prime \cdot \x^\prime}. \label{Psi}
\end{equation}
Equation \rf{Psi} is uniquely solvable if $\psi$ satisfies the radiation
conditions \rf{rc}. From \rf{Psi} it follows
\begin{equation}
(-\Delta -k^2) \psi = -q(\x) \left(2e^{i\bk^\prime \cdot \x^\prime} + \psi
\right)=2ik\gamma(\x^\prime) \left(2e^{i\bk^\prime \cdot \x^\prime} + \psi(\x^\prime)
\right)\delta(x_3). \label{psi1}
\end{equation}
Let us denote by $c(\x^\prime)$ the coefficient of $\delta(\x^\prime)$ in the right hand side of \rf{psi1}
\begin{equation}
c(\x^\prime) = 2ik\gamma(\x^\prime) \left(2e^{i\bk^\prime \cdot \x^\prime} + \psi(\x^\prime)
\right). \label{c}
\end{equation}
In this notation equation \rf{psi1} has the form
\begin{equation}
(-\Delta -k^2) \psi = c(\x^\prime)\delta(x_3). \label{c1}
\end{equation}
Observe that coefficient $c(\x^\prime)$ vanishes outside the support of $\gamma(\x^\prime)$ and
hence solution of equation \rf{c1} can be written as
\begin{equation}
\psi(\x) = \int_{\supp \,\gamma}G(\x - \y^\prime) c(\y^\prime)\,d\y^\prime,
\label{psi2}
\end{equation}
where $\dst G(\x - \y) = \frac{1}{4\pi}\frac{e^{ik |\x - \y|}}{|\x -\y|}$ is the Green's function of \rf{c1}. Form \rf{psi2} and \rf{c} we obtain equation for determining $c(\x^\prime)$
\begin{equation}
c(\x^\prime) +q(\x^\prime)  \int_{\supp \,\gamma}G(\x^\prime - \y^\prime) c(\y^\prime)\,d\y^\prime = -2q(\x^\prime)e^{i\bk^\prime \cdot \x^\prime}. \label{q}
\end{equation}
Finally, it is convenient to introduce a modified potential $v(\x^\prime)$ as
\begin{equation}
v(\x^\prime) = \frac{1}{\pi}\, c(\x^\prime)e^{-i\bk^\prime \cdot \x^\prime}.
\label{v}
\end{equation}
Then \rf{q} becomes
\begin{equation}
v(\x^\prime) 2ik\gamma(\x^\prime)  \int_{\supp \,\gamma}G(\x^\prime - \y^\prime) e^{i\bk^\prime \cdot (\y^\prime -\x^\prime)}v(\y^\prime)\,d\y^\prime = \frac{4ik}{\pi}\,\gamma(\x^\prime). \label{q1}
\end{equation}
Thus, if $v(\x^\prime)$ is known, one can find $\gamma(\x)$ from \rf{q1} from the formula
\begin{equation}
\gamma(\x^\prime) = -\frac{iv(\x^\prime) k^{-1}} {4\pi^{-1} + 2\int_{\supp \,\gamma}G(\x^\prime - \y^\prime) e^{i\bk^\prime \cdot (\y^\prime -\x^\prime)}v(\y^\prime)\,d\y^\prime}, \quad \x^\prime \in \supp \, \gamma. \label{q2}
\end{equation}
In the next section, we will describe a method of determining $v(\x^\prime)$ from the
far field pattern $f(\bk,\ta)$ with fixed $\bk$, and this will complete the solution of the inverse impedance problem.


\section{Calculation of the modified potential}

Equation \rf{psi2} contains Green's function of a shifted argument whose
asymptotic behavior has the form
\begin{equation}
G(\x -\y) = \frac{1}{4\pi}\frac{e^{ik|\x|}}{|\x|}\,e^{-ik\ta \cdot \y}\left[1
+ O\left(\frac{1}{|\x|}\right) \right], \quad |\x| \to \infty.
\end{equation}
Substituting it into \rf{rc} and \rf{psi2}, we obtain the following representation for the far field pattern $f(\bk, \ta)$
\begin{equation}
f(\bk, \ta)
= \frac{1}{4} \int_{\supp \,\gamma} e^{-i(k\ta - \bk)\cdot \y^\prime}v(\y^\prime)\,d\y^\prime,
\label{ffield}
\end{equation}
where $\y^\prime = (y_1,y_2,0)$ and $\ta = \x/|\x|$. Our next goal is to select directions $\ta$ so that the integral \rf{ffield} would represent the Fourier coefficients of function $v(\y^\prime)$.

To this end, we write down the incident vector as $\bk = k(\cos \phi_1, \cos \phi_2, \cos \phi_3)$, while
 ${\bt} = (\cos \theta_1, \cos \theta_2, \cos \theta_3)$.
Then  \rf{ffield} becomes
\begin{equation}
f({\bk, \bt}) = \frac{1}{4}\int_{-1}^{1} \int_{-1}^{1}e^{-ik\left[x(\cos \theta_1 - \cos \phi_1) + y(\cos \theta_2 - \cos \phi_2) \right]}\,v(x,y)\,dxdy.
\label{f3d}
\end{equation}
Expression \rf{f3d} can be associated with the Fourier coefficients of $v(x,y)$ if angles $\theta_1=\theta_{1,n1}$ and $\theta_{2,n2}$ are chosen in such a way that
\begin{align}
k(\cos \theta_{1,n1} - \cos \phi_1) &= \pi n_1, \label{t1}\\
k(\cos \theta_{2,n2} - \cos \phi_2) &= \pi n_2, \label{t2}
\end{align}
where $n_1,n_2 = 0, \pm 1, \pm 2, \ldots$, and
\begin{equation}
\dst -\frac{k}{\pi}\,(1+\cos \phi_i) \leq n_i \leq \frac{k}{\pi}\,(1-\cos \phi_i), \quad i=1,2.
\label{n12}
\end{equation}
For those directions defined by the angles $\theta_{1,n_1}$ and $\theta_{2,n_2}$, the measured far field pattern $f_{n_1, n_2}$ will be the Fourier coefficient in the expansion of the modified potential $v(x,y)$
\begin{equation}
f_{n_1, n_2}  = \frac{1}{4}\int_{-1}^{1} \int_{-1}^{1}e^{-\pi i\left(n_1 x + n_2 y\right)}\,v(x,y)\,dxdy.
\label{fmn}
\end{equation}
Hence, $v(x,y)$ has the following Fourier series representation
\begin{align}
v(x,y) &=\sum_{n_1,n_2} f_{n_1 n_2}\, e^{\pi i\left(n_1 x + n_2 y\right)}.
\label{v1}
\end{align}
Formula \rf{q2} along with \rf{v1} provides the solution of the inverse impedance problem.


\section{Asymptotic solution}

Now we are going to modify the previous approach assuming $k \gg 1$ and using the WKB approximation.
The scattered wave $\psi(\x)$ satisfies the Helmholtz equation in the half space $\R^3_{+} =\{{\x}=(x_1,x_2,x_3), \, x_3 >0\}$
\begin{equation}
-\Delta \psi = k^2 \psi, \quad x_3 > 0,
\label{psi}
\end{equation}
and the boundary condition
\begin{equation}
\left. \rule[4mm]{0mm}{0mm}\psi_{x_3} + ik\gamma({\x^\prime}) \psi ({\x^\prime}) \right|_{x_3 = 0} = -2ik\gamma({\x^\prime})\, e^{i\bk^\prime \cdot \x^\prime }, \quad {\x^\prime} = (x_1,x_2,0).
\label{bc_psi}
\end{equation}
In order to find asymptotic behavior of $\psi (\x)$ for large $k$, we will
use the WKB approximation of $\psi (\x)$ in a neighborhood of support
of $\gamma({\x^\prime})$. We will be looking for expansion of $\psi$
in the form
\begin{equation}
\psi (\x) = e^{i{\bk^{\ast}}\cdot{\x}} \sum_{n=0}^{\infty} \Psi_n (\x)
\left(ik\right)^{-n}, \quad \bk^{\ast} = (k_1,k_2, -k_3).
\label{psi_exp}
\end{equation}
Coefficients  in this expansion can be found explicitly. Substituting \rf{psi_exp} into equation \rf{psi} and equating the coefficients of like powers of $k$, we obtain a recurrence system of differential
equation for $\Psi_n (\x)$
\begin{align}
\widehat{\bk^\ast} \cdot \nabla \Psi_0 \label{psi0} & = 0; \\[2mm]
2\widehat{\bk^\ast} \cdot \nabla \Psi_n + \Delta \Psi_{n-1} &= 0, \;\; n \geq 1,
\end{align}
where $\widehat{\bk^\ast}$ denotes the unit vector in the direction of vector
$\bk^\ast$. From the boundary condition \rf{bc_psi}, one can find the initial
condition for $\Psi_n (\x)$ and thus determine all the coefficients $\Psi_n(\x)$.
In particular, \rf{psi0} implies
\begin{equation}
\Psi_0 (x_1,x_2,x_3) = \Phi_0 \left( x_1 + \frac{k_1 x_3}{k_3}, x_2 + \frac{k_2 x_3}{k_3}\right),
\label{Psi0}
\end{equation}
where $\Phi_0$ is an arbitrary differentiable function.
From \rf{bc_psi} it follows that
\begin{equation}
\Psi_0 (x_1,x_2,0) = -\frac{2\gamma (x_1,x_2)}{\gamma (x_1,x_2)-k_3 k^{-1}},
\label{}
\end{equation}
and hence
\begin{equation}
\Psi_0 (x_1,x_2,x_3) = -\frac{2\gamma (x_1+ k_1 k_3^{-1}x_3,x_2+ k_2 k_3^{-1}x_3)}{\gamma (x_1+ k_1 k_3^{-1}x_3,x_2+ k_2 k_3^{-1}x_3)-k_3 k^{-1}}.
\label{}
\end{equation}
Thus, using expansion \rf{psi_exp} and relations \rf{c} and \rf{v}, we obtain the following asymptotic representation for the scattering
amplitude $f$ \rf{ffield} through the boundary impedance $\gamma$
\begin{equation}
f(\bk, \ta) = -\frac{ik_3}{\pi} \int_{\supp \,\gamma}
e^{-i(k\ta - \bk)\cdot \y^\prime}\frac{\gamma(\y^\prime)}{\gamma(\y^\prime) -k_3 k^{-1}}d\y^\prime + O\left( k^{-1} \right).
\label{fasy}
\end{equation}
If we choose the direction of measurements $\ta$ of the far field pattern the same as before in \rf{t1}-\rf{n12}, then the value $f(\bk, \ta)$ becomes proportional to the Fourier coefficient of $\gamma(\gamma -k_3k^{-1})^{-1}$
\begin{equation}
f_{n_1,n_2} = -\frac{ik_3}{\pi} \int_{-1}^{1}\int_{-1}^{1}
e^{-\pi i(n_1y_1 + n_2y_2)}\,\frac{\gamma(y_1,y_2)}{\gamma(y_1,y_2)-k_3 k^{-1}}\,dy_1 dy_2 + O\left( k^{-1} \right).
\label{}
\end{equation}
Applying the inverse Fourier transform, we obtain
\begin{equation}
\frac{\gamma(x_1,x_2)}{\gamma(x_1,x_2)-k_3 k^{-1}} = \frac{\pi i}{4k_3}\sum_{m,n} f_{m,n}\,e^{\pi i (mx_1 + nx_2)}+ O\left( k^{-1} \right).
\label{gf}
\end{equation}
This equation can be solved for $\gamma$. Hence the boundary impedance can be restored using the values of the far field pattern in the specific directions given by \rf{t1}-\rf{t2}.

\section{Scattering from sphere}


In the case of convex body $\Omega$,  there is a direct asymptotic relation between the far field pattern and the boundary impedance for large values of $k$ \cite{M76}
\begin{equation}
\gamma ({\B y}^+) = \frac{\K^{\;\;-\frac{1}{2}}({\B y}^+)+2|f({\bk}, \ta)|}{\K^{\;\;-\frac{1}{2}}({\B y}^+)-2|f({\bk}, \ta)|}\; {\B n}\cdot \ta + O(k^{-1}),
\label{ma}
\end{equation}
where ${\B y}^+ (\ta) \in \partial \Omega$ is the preimage of ${\B n} = (\ta - \widehat{\B k})/|\ta - \widehat{\B k}|$ under the Gauss map, and $\K\; ({\B y}^+)$ is the Gauss curvature at ${\B y}^+ \in \Omega$.
Formula \rf{ma} has asymptotic character, and we want to figure out the range of values of $k$ that give a good approximation
of $\gamma$. We also analyze the approximation of the far field pattern $f({\bk}, \ta)$ by the measurement
of the scattered field at the distance $r$.

In order to conduct a numerical experiment, we restrict ourselves to the case where the direct problem can be easily solved. For that purpose we consider the problem that has an exact solution -- scattering of plane wave $e^{ikz}$ from a sphere of radius $a$ with constant boundary impedance. Then the formula \rf{ma} takes the form
\begin{equation}
\gamma = \frac{a+2|f({\bk}, \ta)|}{a-2|f({\bk}, \ta)|}\sin \frac{\theta}{2} + O(k^{-1}), \quad \frac{\pi}{2} \leq \theta \leq \pi,
\label{gam_asy}
\end{equation}
where $\theta$ is the polar angle on sphere.

Similar to \rf{r3}, we need to solve the boundary value problem for the Helmholtz equation
\begin{align}
&-\Delta u = k^2 u, \quad r > a, \label{sp} \\[2mm]
&\left. u_{r} - ik\gamma\, u \right|_{r = a} = 0,  \label{bc_sp}
\end{align}
where $\gamma>0$ is a constant surface impedance and
\begin{equation}
u({\x}) = e^{ikz}+ \phi({\x})
\label{sp1}
\end{equation}
with $\phi({\x})$ satisfying the radiation condition \rf{rc}.

Solution of the problem \rf{sp}-\rf{sp1} is given by
\begin{equation}
u = e^{ikz} - \sum_{n=0}^{\infty}(2n+1)i^n \frac{nj_{n-1}(ka)-(n+1)j_{n+1}(ka)-i\gamma j_n (ka)}{nh_{n-1}(ka)-(n+1)h_{n+1}(ka)-i\gamma h_n (ka)}\, h^{(1)}_{n}(kr) P_{n}(\cos \theta),
\label{sp_bvp}
\end{equation}
where $j_{n}(z)$ and $h^{(1)}_{n}(z)$ are spherical Bessel functions of the first and third kind, respectively, and $P_{n}(x)$ are Legendre polynomials \cite{AS65}. From this formula we can determine the far field pattern $f({\bk}, \ta)$ \rf{rc} and calculate the surface impedance $\gamma$ using it asymptotics \rf{gam_asy} for large values of $k$.

\section{Numerical examples}

\begin{figure}[hbt]
\centering
\includegraphics[width=0.45\linewidth, angle=0]{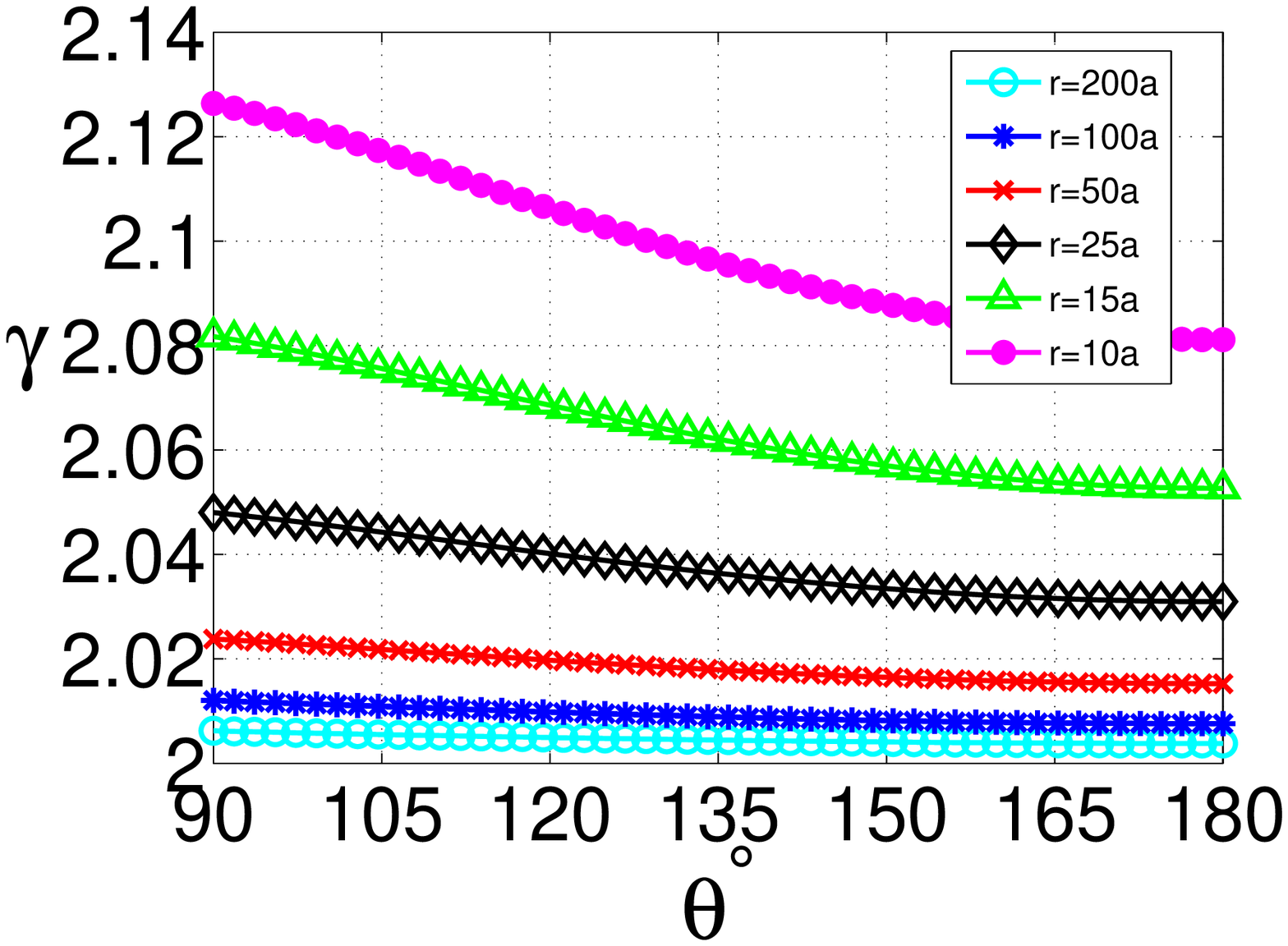} \hspace{0mm}
\includegraphics[width=0.45\linewidth, angle=0]{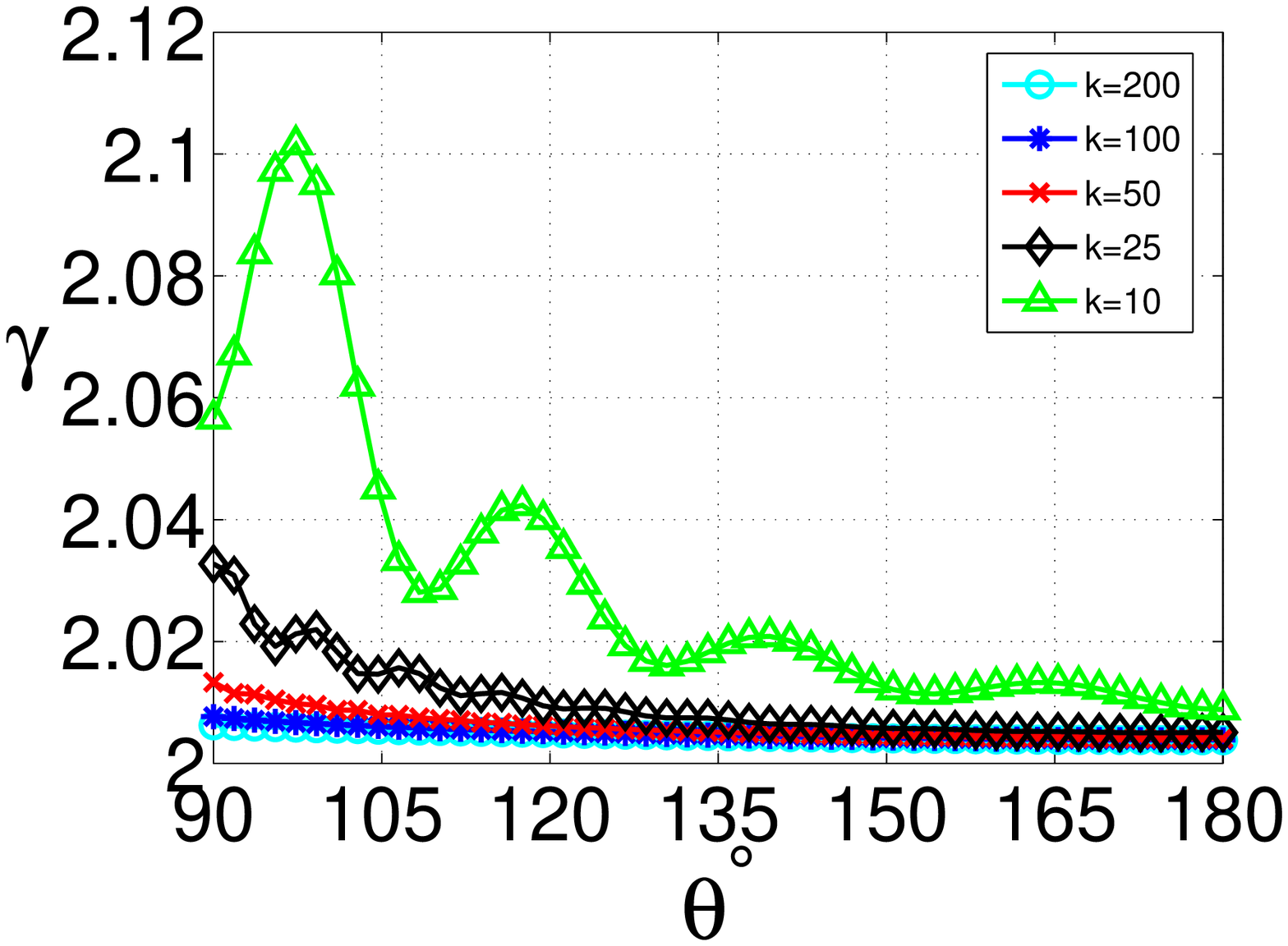}\\
\hspace{5mm}(a)\hspace{70mm}(b)
\caption{Reconstruction of the boundary impedance $\gamma = 2$ of the illuminated part of the sphere of radius $a=1$ from the formula \rf{gam_asy}. Left: the far field pattern $f(k,\ta)$ with the wave number $k=200$ is approximated by the amplitude of the scattered wave at different distances $r$ from the center of the sphere. Right: the far field pattern $f(k,\ta)$ is approximated as before with $r=200a$,  wave numbers $k$ vary. Deviation from $\gamma = 2$ increases as angle $\theta$ approached $90^\circ$ where incident rays are tangent to the sphere.
}


\label{fig1}
\end{figure}

Figure \ref{fig1} shows the reconstructed boundary impedance $\gamma = 2$ of a unit sphere based on the asymptotic formula \rf{gam_asy}. In the left figure, the far field pattern with the wave number $k=200$ was determined from the exact solution using \rf{rc}
\begin{equation}
|f({\bk}, \ta)| \approx |\x | \psi (\x )
\end{equation}
for various distances $|\x |$ from the sphere. The accuracy of approximation monotonically improves as the polar angle $\theta$ increases and does not exceed about $0.5\%$ for the distances beyond $r=100a$. The right figure shows the dependence of the restored impedance on the wave number of the incident plane wave while at the distance $r=200a$ from the sphere. As the wave number $k$ decreases, not only approximation of $\gamma$ deteriorates, but it also starts to exhibit oscillatory behavior.  Approximation of $\gamma$ is also improving for larger values of $\theta$ and remains below $0.5\%$ as long as $k > 50$.

\begin{figure}[hbt]
\begin{center}
\includegraphics[width=0.45\linewidth, angle=0]{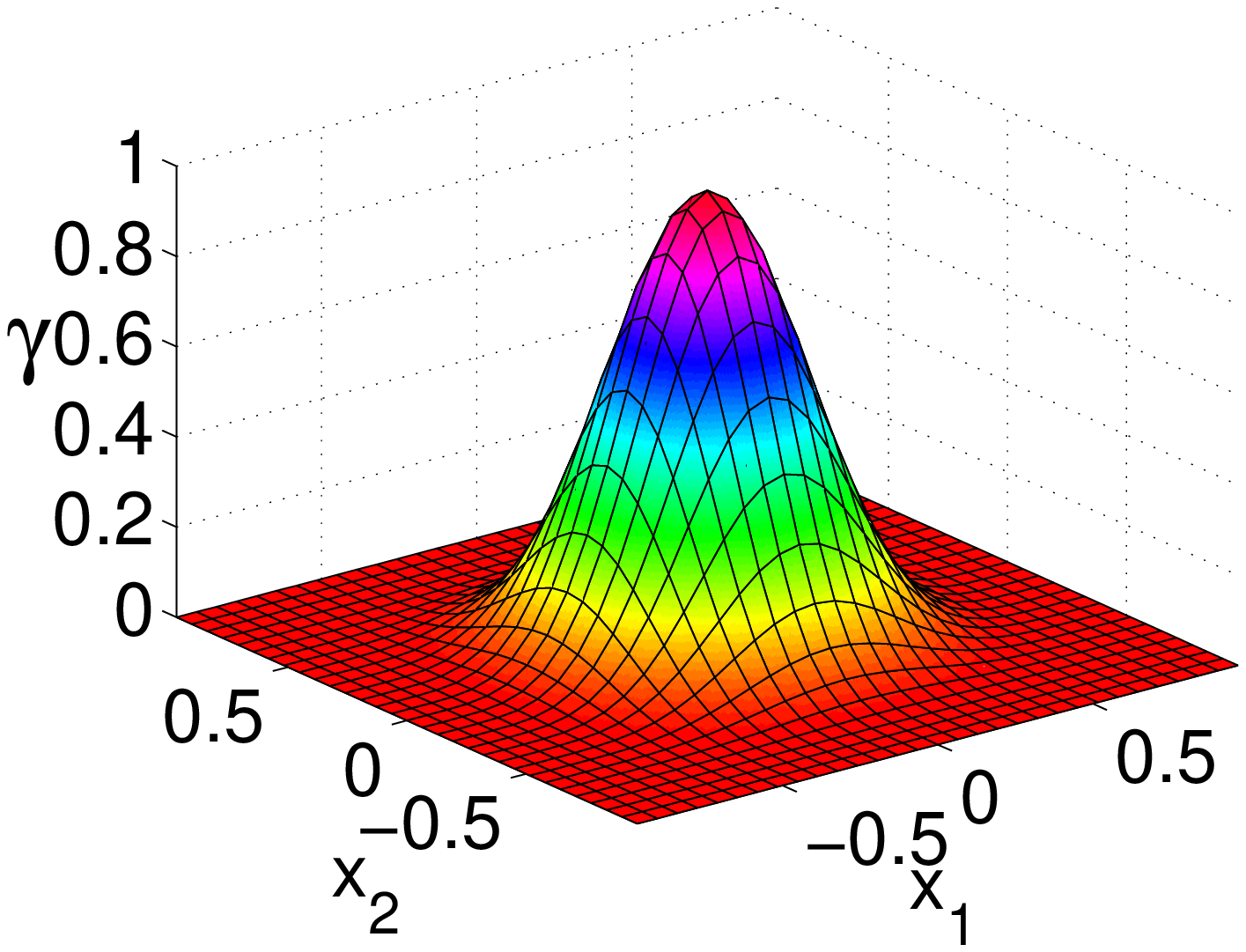} \hspace{0mm}
\includegraphics[width=0.45\linewidth, angle=0]{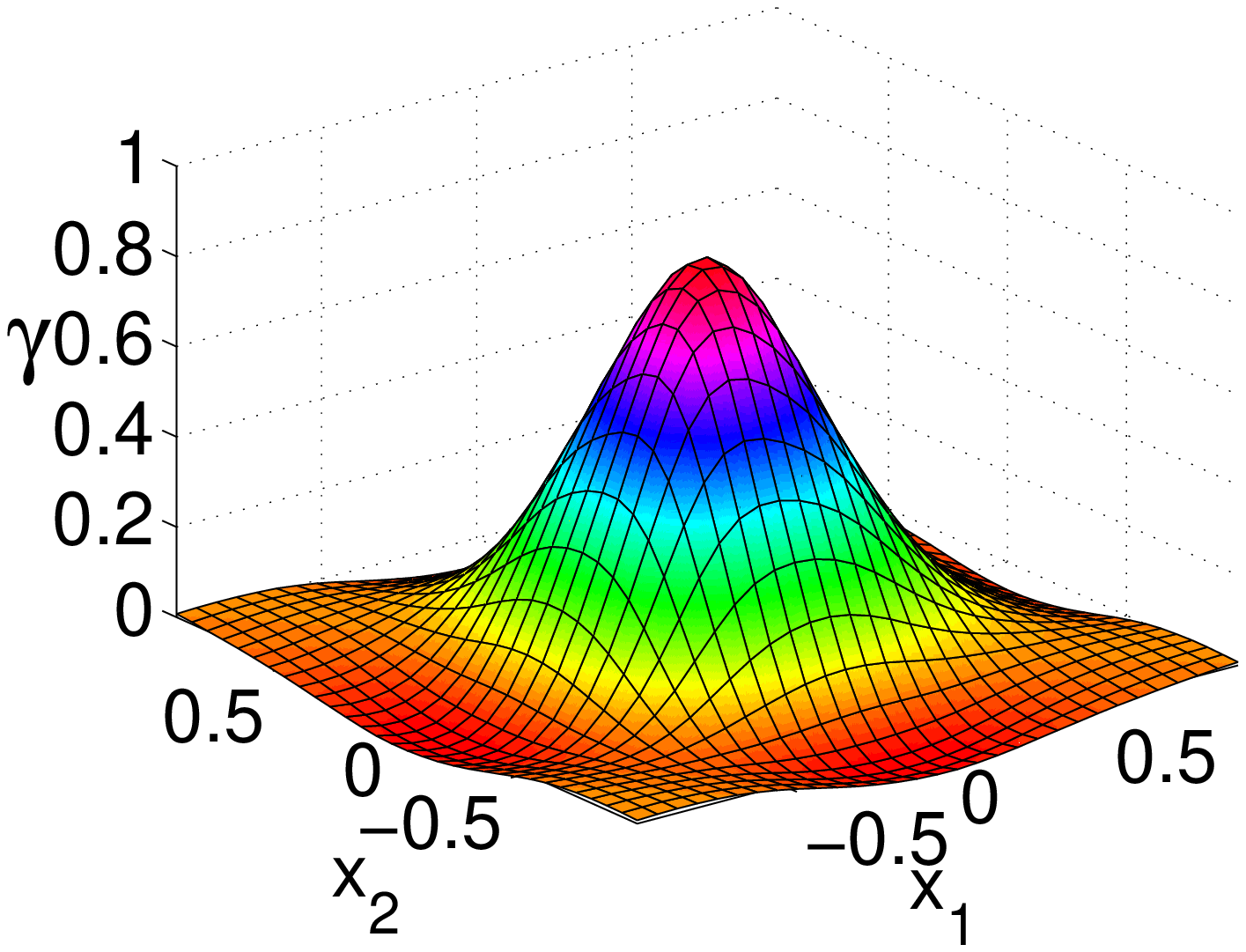}\\
\hspace{0mm}(a)\hspace{75mm}(b) \\
\includegraphics[width=0.45\linewidth, angle=0]{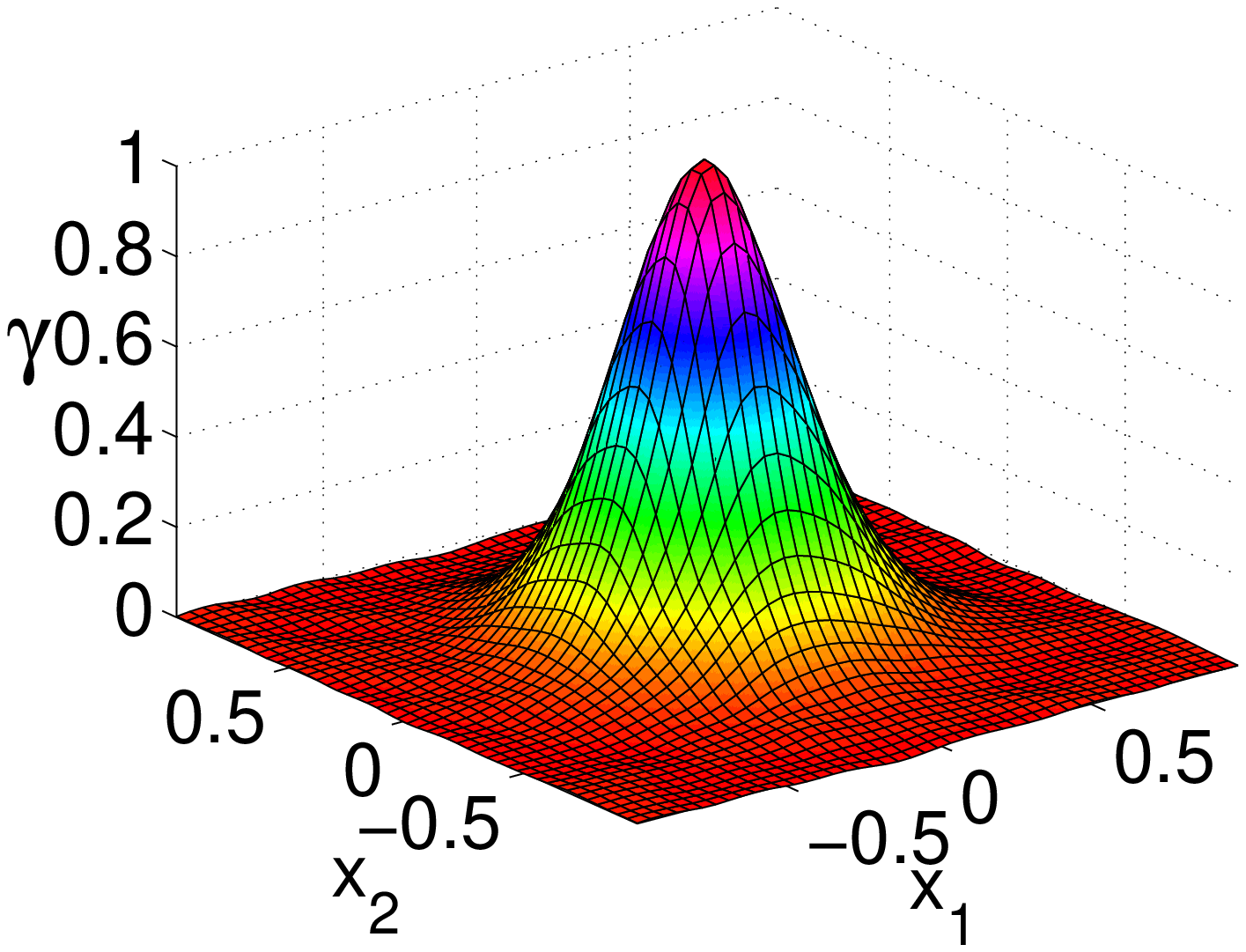}
\hspace{0mm}
\includegraphics[width=0.45\linewidth, angle=0]{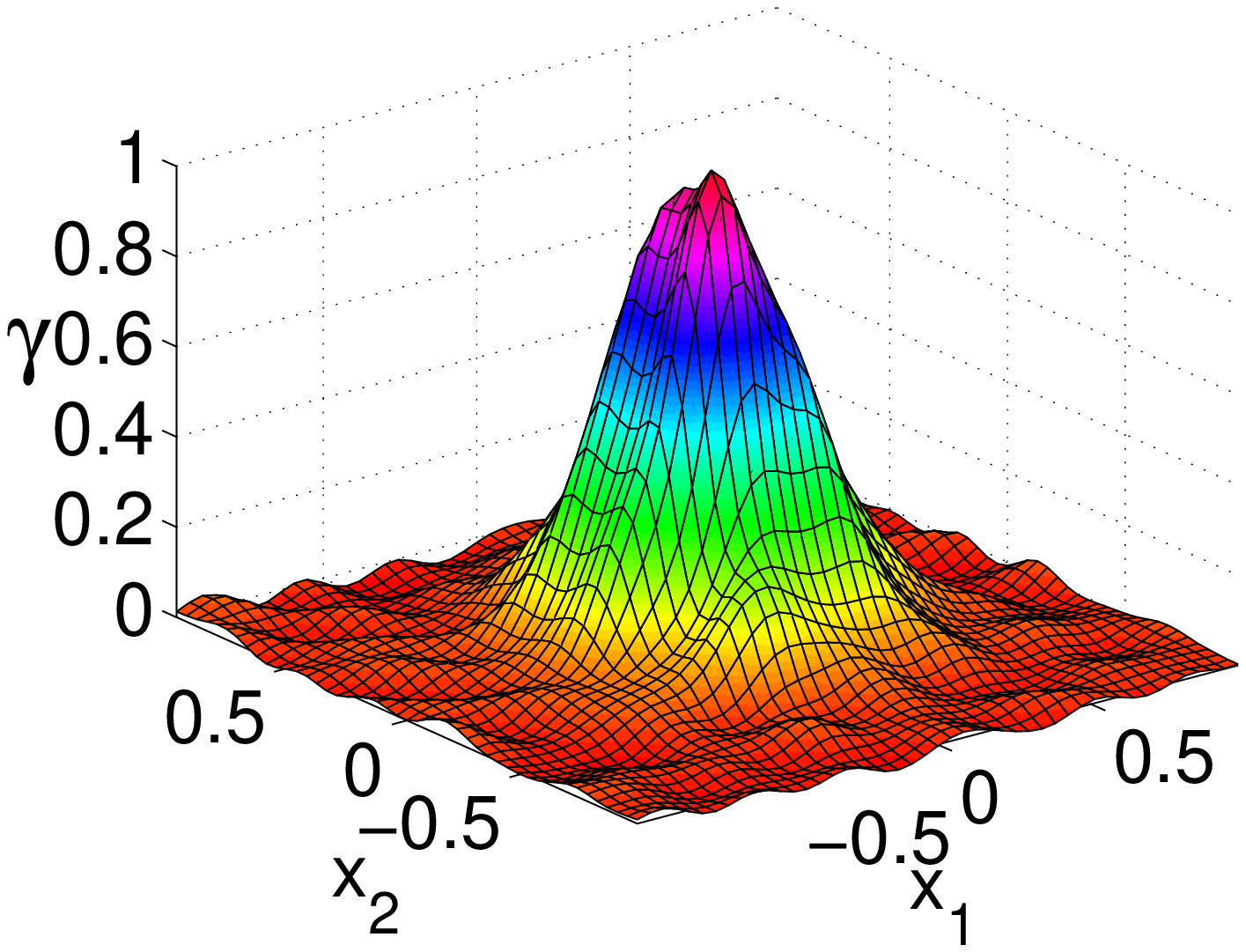}\\
\hspace{0mm}(c)\hspace{75mm}(d)
\end{center}
\caption{Reconstruction of the boundary impedance $\gamma(x)$ (a) using explicit formula formula \rf{q2} for different values of the wave vector $k$. In (b) $k=10$. $k=15$ with $1\%$ (c) and $5\%$ (d) additive random  noise.}
\label{fig2}
\end{figure}

The above approach leads to a good  approximation of the impedance if the far field is measured at a distance by order of magnitude greater than the diameter of the sphere and for the wave length that is by order of magnitude lesser than the diameter of the sphere. Similar results are observed in restoring a compactly supported boundary impedance of a half space. Using the measurements of the far field pattern and both formulas  \rf{gf}, \rf{q2}, and \rf{v}, we reconstructed the boundary impedance \ref{fig2}(a).  In figure \ref{fig2}, we used explicit formula \rf{q2} with $k=10$ in (b). Then the Fourier coefficients of the far field pattern were perturbed by random numbers uniformly distributed in the interval $[-1,1]$. Figures \ref{fig2}(c)-(d) show reconstructed boundary impedance for $k=15$ when the amplitudes of the additive random noise were $1\%$ and $5\%$ of the greatest Fourier coefficient, respectively.

\begin{figure}[hbt]
\begin{center}
\includegraphics[width=0.45\linewidth, angle=0]{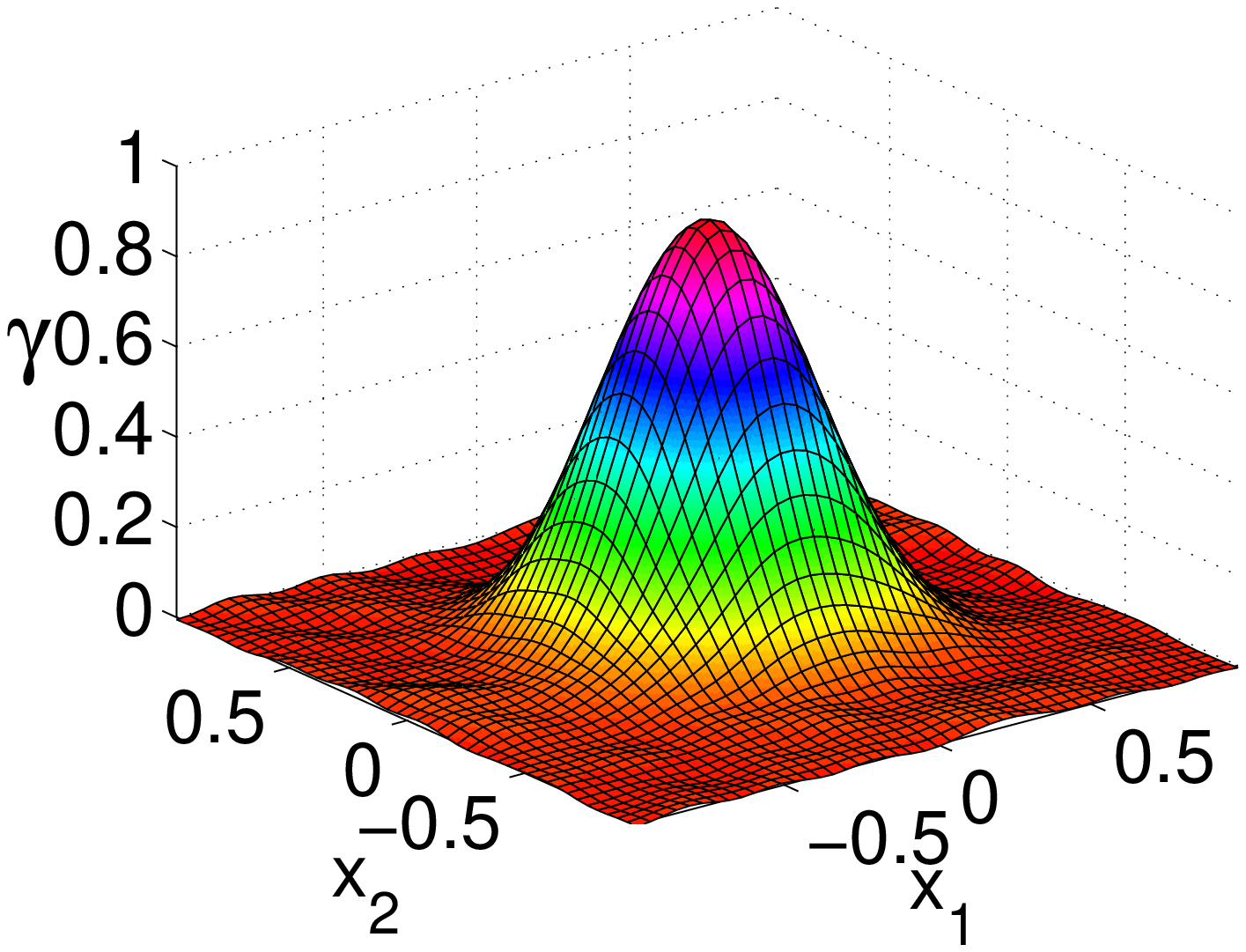}
\hspace{0mm}
\includegraphics[width=0.45\linewidth, angle=0]{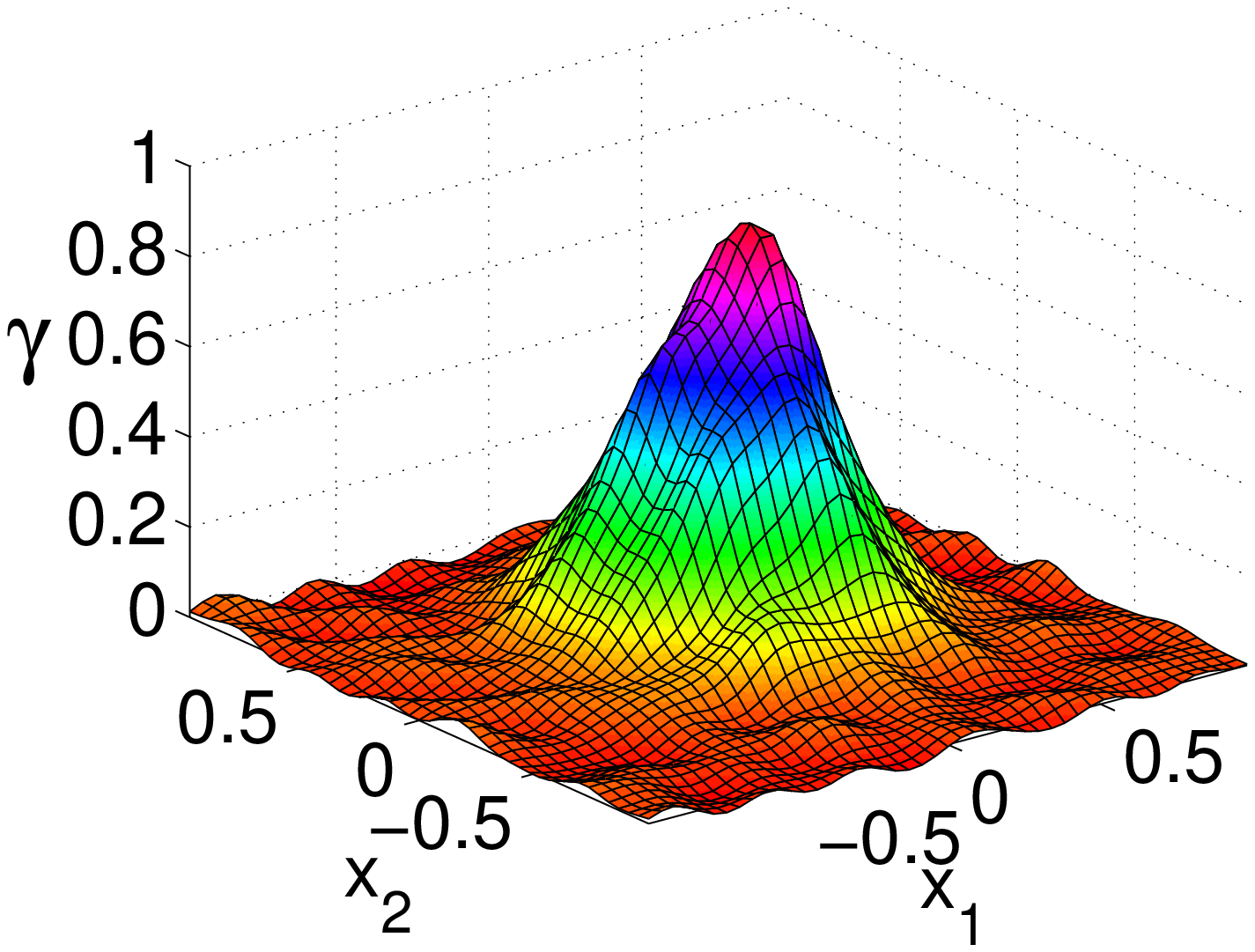}\\
\hspace{0mm}(a)\hspace{75mm}(b)
\end{center}
\caption{Reconstruction of the same boundary impedance $\gamma(x)$ from figure \ref{fig2}a using asymptotic formula \rf{gf} when $k=15$ and the amplitude of the additive random  noise is $1\%$ (a) and $5\%$ (b), respectively.}
\label{fig3}
\end{figure}

Reconstruction of the boundary impedance by asymptotic formula \rf{gf} gives slightly lesser accuracy as compared with exact formula \rf{q2}. 
In figure \ref{fig3} we reconstructed the boundary impedance from figure \ref{fig2}a using asymptotic formula \rf{gf} and $k=15$. The Fourier coefficients of the far field pattern are corrupted by an additive uniformly distributed random noise form the interval $[-1,1]$ with the amplitude $1\%$ (a) and $5\%$ (b) of the largest Fourier coefficient. The accuracy of reconstruction in this case is slightly less as compared with exact formula \rf{q2}.

\begin{figure}[hbt]
\begin{center}
\includegraphics[width=0.45\linewidth, angle=0]{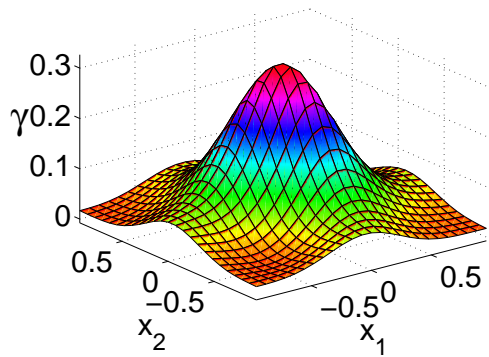}
\end{center}
\caption{Reconstructed boundary impedance $\gamma(x)$ from figure \ref{fig2}a using either explicit \rf{q2} or asymptotic formula \rf{gf} when $k=5$ and the amplitude of the additive random  noise is $1\%$.}
\label{fig4}
\end{figure}

Finally, in figure \ref{fig4} we reconstruct the impedance in the presence of a $1\%$ additive random noise when the wavenumber is as small as $k=5$. Although in this case there is a significant error in the restored amplitude, it captures qualitatively the shape and the location of the inhomogeneity of the impedance.

\section{Conclusions}

We have considered the problem of determining a compactly supported boundary impedance from knowledge of the time harmonic incident wave and its far field pattern. The approach is based on a special selection of the directions in which the far field pattern is measured. Then the boundary impedance is expressed through a potential using a simple exact formula, while the Fourier coefficients of the potential equal the measured far field patterns. Efficiency of the approach is illustrated by numerical examples.

%
%
%
%
%
%

\end{document}